\documentclass[epj]{svjour}
\usepackage{graphics}

\begin{document}

\title{Wetting and interfacial adsorption in the Blume-Capel model on the square lattice}

\author{N.G. Fytas\inst{1} \and W. Selke\inst{2}
}                     

\institute{Applied Mathematics Research Centre, Coventry
University, Coventry, CV1 5FB, United Kingdom \and Institut
f\"{u}r Theoretische Physik and JARA-HPC, RWTH Aachen University, 52056 Aachen,
Germany}

\date{Received: date / Revised version: date}

\abstract{We study the Blume-Capel model on the square lattice. To
allow for wetting and interfacial adsorption, the spins on
opposite boundaries are fixed in two different states, ``+1'' and
``-1'', with reduced couplings at one of the boundaries. Using
mainly Monte Carlo techniques, of Metropolis and Wang-Landau type,
phase diagrams showing bulk and wetting transitions are
determined. The role of the non-boundary state, ``0'', adsorbed
preferably at the interface between ``-1'' and ``+1'' rich regions, is
elucidated.
}
\authorrunning{N.G. Fytas and W. Selke} \titlerunning{Wetting and interfacial adsorption in the Blume-Capel model on the square lattice}

\maketitle

\section{Introduction}
\label{sec:intro}

Critical interfacial phenomena have been studied extensively in
the last decades, both experimentally and
theoretically~\cite{Abra,Diet,Bonn,Ral}. A well-known example is
wetting, where the macroscopically thick phase, e.g., the fluid,
is formed between the substrate and  the other phase,
say, the gas. Liquid and gas are separated by the interface. The
scenario may be mimicked, in Statistical Physics,  in a simple way
by the two-state Ising model, with the state ``+1'' representing,
say, the fluid, and ``-1'' the gas.

An interesting complication arises when one considers the
possibility of more than two phases. A third phase may be formed
at the interface between the two other
phases. An experimental realization is the two-component fluid
system in equilibrium with its vapor phase~\cite{Diet,Mold}. The
situation may be mimicked in a simplified fashion in three-state
models, like Potts~\cite{Pesch} or Blume-Capel
models~\cite{Kroll}. The formation of the third phase in such
models has been called ``interfacial adsorption''.

In the following article, we shall consider wetting in the
(3-state) Blume-Capel model, with spin $S=1$, on the square
lattice, paying special attention to possible interfacial
adsorption. The present analysis has been motivated by the closely
related recent Monte Carlo (MC) study~\cite{Alba}. In that
thorough study, wetting had been imposed by surface
fields of symmetric strength but opposite sign. Here, spins at
opposite boundaries are fixed in two
different states, reducing, in addition, the couplings to one of
the boundaries. Thence, there will be no ambiguity in the wetting
phenomenon occurring only relative to the boundary with reduced
couplings. Similar boundary conditions  have been used before for
describing interfacial phenomena in Ising~\cite{Doug} and
Potts~\cite{Pesch2} models.

The outline of the article is as follows: In the next
section~\ref{sec:mm}, the model and the methods, especially, MC
simulations of Metropolis and Wang-Landau type, will be
introduced, followed by the discussion of our main results in
section~\ref{sec:results}. The summary, section~\ref{sec:summary},
will conclude the article.

\section{Model and Methods}
\label{sec:mm}

The Blume-Capel (BC) model on the square lattice is a classical
spin-1 Ising model described by the Hamiltonian~\cite{Blume,Capel}
\begin{equation}
\label{eq:Hamiltonian} \mathcal{H}=-J\sum_{\langle (i,j)
\rangle}S_{i,j}S_{i\pm1,j\pm1}+D\sum_{i,j}S_{i,j}^{2},
\end{equation}
where the spin variable $S_{i,j}$, at site $(i,j)$, takes
on the values -1, 0, or +1, with the sums running over the
entire lattice. The exchange interaction between neighboring
spins is ferromagnetic
$J>0$; $D$ denotes the strength of the single-ion anisotropy term,
where $D\ge 0$.

The bulk phase diagram of the model has been studied extensively
and carefully, using approximate, but highly accurate
methods~\cite{nightingale82,beale86,silva06,malakisbc}.
From these analyzes, the BC model is known to order, at low
temperatures, ferromagnetically for $D<2$. There is no phase
transition for $D>2$, where $S_{i,j}=0$ at each site in the ground
state. The transition temperature, $k_BT_c/J$, decreases
monotonically with increasing value of $D$, approaching zero in
the limit $D=2$. The phase transition is continuous at $D<D_t$,
whereas it becomes of first-order at $D>D_t$. At $D_t=1.9655(10)$
and $k_BT_t/J$=0.610(5)~\cite{beale86}, a tricritical point
occurs. Note that this rather old estimate has been confirmed more
recently \cite{silva06}.

\begin{figure}
\resizebox{1 \columnwidth}{!}{\includegraphics{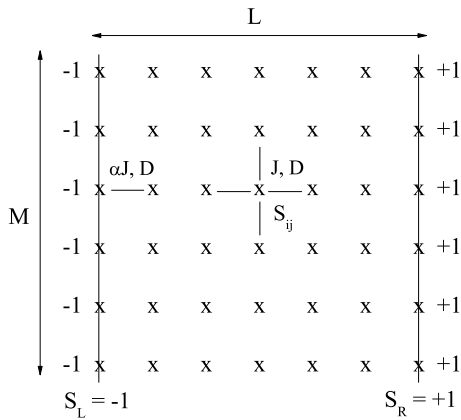}}
\caption{Geometry and interactions of the model.}
\label{bcfiggeometry}
\end{figure}

To study wetting, we shall employ special boundary conditions,
modifying, in addition, the exchange interaction at one of the
boundaries: As sketched in figure~\ref{bcfiggeometry}, the spins
of the lattice with $L \times M$ sites are fixed on one of the
boundaries, say, the left boundary, in state $S_L=-1$, while the
spins on the opposite boundary are fixed to be $S_R=1$. The two
boundaries are separated by $L$ sites. Top and bottom  boundaries
are connected by periodic boundary conditions. Obviously, then an
interface between regions of predominantly -1 and +1 spins may be
formed. To allow for wetting or (de)pinning of the interface, the
couplings at one of the boundaries may be modified, for example by
introducing, at the left hand side, the surface coupling $\alpha
J$ between the boundary spins and the neighboring bulk spins, with
$0 \le \alpha \le 1$. Otherwise, the couplings between neighbors
are always $J$ \cite{Doug}. Wetting may then take place when the ``-1'' rich
region spreads from the left hand side towards the center of the
lattice. In the limiting case $\alpha=0$, i.e. assuming free
boundary condition at the left boundary, there will be, of course,
no wetting. When $\alpha=1$, the interface will be de-pinned at
arbitrarily low temperatures. Interfacial adsorption will show up
with spins in the state ``0'' being preferably adsorbed at the
interface between ``-1'' and ``+1'' rich
regions~\cite{Pesch,Kroll}. We shall mainly deal with the case $0
< \alpha < 1$, where both wetting and interfacial adsorption
may play an interesting role.

To analyze the model, studying both critical bulk and wetting
phenomena, we used large-scale MC simulations, augmented by rather
straightforward energy considerations for the ground state, $T=0$,
and for low temperature excitations.

In particular, to determine accurately the bulk transition
temperatures for our special choices of the strength of the
single-ion anisotropy, $D$, we implemented a modified
version~\cite{malakisWL,fytasWL1,fytasWL2} of the Wang-Landau (WL)
algorithm~\cite{wang}. The modifications have been proposed to
cope with the huge number of energy states for large lattices
and to deal with the important aspect of detailed balance. More
concretely, a combination of several stages of
the WL process has been suggested. In short, we carry out a
starting ''multi-range
(multi-R) stage'', in a very wide energy interval, by subdividing that energy
interval in several subintervals overlapping at one or
several points \cite{malakisWL}, and up to a certain
level of the WL random walk. The WL refinement is $G(E)\rightarrow
f G(E)$, where $G(E)$ is the density of states (DOS), and we
follow the usual modification factor adjustment
$f_{j+1}=\sqrt{f_{j}}$ and $f_{1}=e$. This preliminary stage
consists of the levels $j=1,\ldots,18$ and to improve accuracy the
process is repeated several times. The process continues in two
further stages, using now the high WL iteration levels, where the
modification factor is very close to unity and there is not any
significant violation of the detailed balance condition. In the
first (high-level) stage, we follow again a repeated several times
(typically $\sim 5-10$) multi-R WL approach, carried out now in a
restricted energy subspace obtained form the preliminary stage,
following the prescription of
references~\cite{malakisWL,fytasWL1}. The WL levels may be now
chosen as $j=18,19,20$ and as an appropriate starting DOS for the
corresponding starting level the average DOS of the preliminary
stage at the starting level may be used. Finally, the second
(high-level) stage is applied in the refinement WL levels
$j=j_{i},\ldots,j_{i}+3$ (typically $j_{i}=21$), where we use an
one-range approach, i.e. for a fixed energy interval, where the
adjustment of the WL
modification factor follows the rule $\ln f\sim t^{-1}$, where $t$
the MC time~\cite{belardinelli07}.

On the other hand, to investigate the wetting transition, we
applied the simpler standard Metropolis MC algorithm, as
had been done before~\cite{Alba}. We still achieved a good
accuracy, performing runs of sufficient length, as we checked, for
instance, by comparing WL and Metropolis data for the specific heat and
magnetic susceptibility.

\section{Results}
\label{sec:results}

\subsection{Phase diagram}
\label{sec:phadia}

The phase diagram of the BC model on the square lattice comprises
the bulk phase transition, $k_BT_c/J$, and, usually at lower
temperature, the wetting transition, $k_B T_w/J$.

\begin{figure}
\resizebox{1 \columnwidth}{!}{\includegraphics{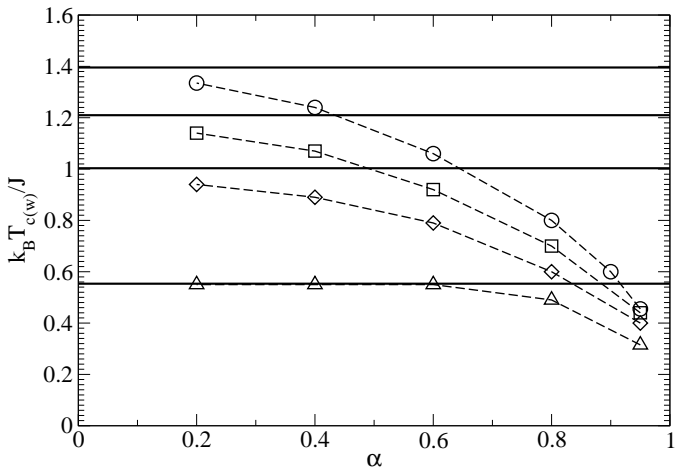}}
\caption{Bulk phase transitions, $k_BT_c/J$, solid lines, and
wetting transitions, $k_BT_w/J$, versus reduction factor $\alpha$
for $D/J=1.0$ (circles), $1.4$ (squares), $1.7$ (diamonds), and
$1.98$ (triangles); dashed lines are guides to the eye.}
\label{bcfigphadia}
\end{figure}

Of course, $T_c$ depends only on the strength of the single-ion
anisotropy term, $D/J$,  but not on the reduction factor of the
boundary couplings, $\alpha$. As depicted in
figure~\ref{bcfigphadia}, we studied the cases $D/J=1.0$, $1.4$,
$1.7$, and $1.98$, with $\alpha$ ranging from $0.2$ to $0.95$,
using MC techniques.

For continuous bulk transitions, $D/J < D_t/J$, the wetting line
is found to decrease monotonically with increasing $\alpha$, where
$T_{w}(\alpha=0)=T_c$ and $T_{w}(\alpha=1)=0$. Interestingly, for
bulk transitions of first-order, $D/J=1.98$, the wetting line
$T_{w}(\alpha)$ ends, at non-vanishing value of $\alpha$, in the
bulk transition, as has been observed in the two-dimensional BC
model for wetting induced by surface fields~\cite{Alba}.

In the following, we shall discuss these main findings in detail.

\subsection{Bulk transition}
\label{sec:bulk}

The square-lattice BC model, equation~(\ref{eq:Hamiltonian}), at
the crystal-field values $D/J < D_t/J$ undergoes a second-order
phase transition between the ferromagnetic and paramagnetic
phases, expected to be in the universality class of the
two-dimensional Ising model~\cite{beale86}. This aspect
had been verified recently by high-accuracy WL type of simulations
for single-ion anisotropy values in the range
$D/J=[0-1.8]$~\cite{silva06,malakisbc}.

Here we study the bulk transition in the second-order regime
of the phase diagram for three values
of $D/J$, namely $D/J=1.0$, $1.4$, and $1.7$. Using the
above sketched WL MC approach we simulated quadratic systems with periodic
boundary conditions and linear sizes $L$ ranging from $20$ to
$100$. Moreover, to increase statistical accuracy, we averaged,
for each pair $(L,D/J)$, over at least $50$ independent runs.

The case $D/J=1.0$ has been investigated
before~\cite{beale86,malakisbc}, and so a direct comparison is
possible. For the other two cases, we present new
estimates for the critical temperatures. As we shall see below,
they agree well with the, presumably, best previous estimates for
nearby values of the single-ion anisotropy~\cite{beale86}.

\begin{figure}
\resizebox{1 \columnwidth}{!}{\includegraphics{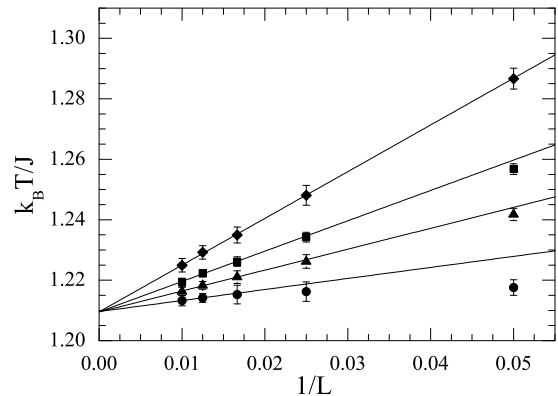}}
\caption{Peak positions, for $D/J=1.4$, of the specific heat
(circles), susceptibility (squares), logarithmic derivative of the
absolute order parameter (triangles), and logarithmic derivative
of the second moment of the magnetization (diamonds). The common
extrapolation to the thermodynamic limit, $T_c$, is shown by the
solid lines.} \label{bcfigTc}
\end{figure}

For each value of $D/J$, we performed a standard finite-size
scaling analysis in order to estimate the critical temperature
$T_c$. For reasons of brevity, we show here only the typical case
of $D/J=1.4$ in figure~\ref{bcfigTc} below (similar analysis has
been performed for the other two cases). In the following we
use the abbreviation $K=J/(k_BT)$ for the inverse temperature. $E$ denotes the
energy, and $m=(1/N)\sum_{1}^{N}S_{i,j}$ is the magnetization, with
the number of lattice sites $N=L^{2}$. In figure~\ref{bcfigTc} we
plot the shift-behavior of different ``pseudo-critical''
temperatures corresponding to the peak positions
of four quantities: The specific heat
\begin{equation}
\label{eq:C} C=K^{2}N^{-1}\left[\langle E^2 \rangle - \langle E
\rangle^2\right],
\end{equation}
the magnetic susceptibility
\begin{equation}
\label{eq:chi} \chi=KN\left[\langle m^2 \rangle - \langle |m|
\rangle^2\right],
\end{equation}
the derivative of the absolute value of the magnetization
\begin{equation}
\label{eq:absm} \frac{\partial \langle |m|\rangle}{\partial
K}=\langle |m|\mathcal{H}\rangle-\langle |m|\rangle\langle
\mathcal{H}\rangle,
\end{equation}
and the logarithmic derivative of the second moment of the
magnetization
\begin{equation}
\label{eq:m2} \frac{\partial \ln \langle m^{2}\rangle}{\partial
K}=\frac{\langle m^{2}\mathcal{H}\rangle}{\langle
m^{2}\rangle}-\langle \mathcal{H}\rangle.
\end{equation}
Fitting simultaneously our data for the larger lattice sizes to
the expected asymptotic power-law behavior $T=T_{c}+bL^{-1/\nu}$, and taking,
by invoking Ising universality, $\nu=1$, we obtain the following
estimates for the critical temperatures $k_BT_c/J$=1.3957(7),
1.2098(8), and 1.0033(7) at $D/J$= 1.0, 1.4, and 1.7, respectively. The
error bars take into account the dependence of the estimates for
$T_c$ on the system sizes included in the fits, having in mind
the possible effect of corrections to scaling.

We now turn to the case $D/J=1.98 > D_{t}/J=1.9965$, where the
bulk transition is of first-order. There, in accordance with the
finite-size scaling theory of first-order phase
transitions~\cite{fisher82,challa86,binder87}, the MC data for
the peak positions may be
fitted according to the double Gaussian approximation. We
then obtain for the bulk transition
temperature $k_BT_c/J=T^{\ast}=0.5533(9)$. This finding is corroborated by an
analysis of the distribution function of the energy density, with
two distinct peaks of equal height at the
transition~\cite{binder87,lee90,janke93}.

Our estimates for the bulk transition temperatures are marked by
the solid lines in figure~\ref{bcfigphadia}, where the thickness
of the lines is larger than the error bars we stated.

Closing this subsection on the bulk transition of the BC model for
the four distinct values of $D/J$, a few comments may be added.
Regarding the accuracy of our final estimates: Our
estimate for $T_c$  at $D/J=1.0$ agrees well with the previous
estimate, $1.398(2)$ of~\cite{malakisbc}, and the estimates for
$D/J=1.4$ and $1.7$ are fully compatible with those given in Table
I of reference~\cite{beale86} for nearby values of $D/J$. Also, in
the first-order transition regime, the value of the transition
temperature for $D/J=1.98$ interpolates reasonably well between the
estimates $T^{\ast}(D/J=1.969)=0.60$ and $T^{\ast}(D/J=1.99)=0.55$
of~\cite{beale86} (see also~\cite{silva06}). Regarding the type of
phase transitions: As expected, the transitions of second order are
observed to be in the universality class of the two-dimensional
Ising model, with, e.g., the maximal susceptibility growing with
the exponent $7/4$. For the first-order transition, we
confirmed the expected finite-size scaling behavior, e.g., of the maximal
susceptibility, $\sim L^{2}$, as seen in our WL data for
$D/J=1.98$.

\subsection{Wetting}
\label{sec:wetting}

To study wetting in the BC model, we mainly performed Metropolis
MC simulations, augmented by rather straightforward ground-state
and low-temperature energy considerations.

At $T=0$, for $0<\alpha<1$ either all spins are in the state
``+1'' or, if
\begin{equation}
\label{eq:pw}  D/J \ge 2 - \alpha ,
\end{equation}
there is a single line of
spins in the state ``0'' next to the left hand boundary, see
figure~\ref{bcfiggeometry}. This ``pre-wetting'' phenomenon had
been observed before for $\alpha=1$ ~\cite{kote}.
At non-zero temperatures, the pre-wetting may be enhanced, or
reduced. Indeed, if $D/J \ge 2-(\alpha/2)$, it costs less energy
to flip a ``+'' spin next to the line of ``0''s to the state ``0''
than to do the reverse flip for a spin in the pre-wetting line. In
any event, the possible effect of pre-wetting by ``0''s on the
critical wetting, at $T_w > 0$, will be discussed below.

\begin{figure}
\resizebox{1 \columnwidth}{!}{\includegraphics{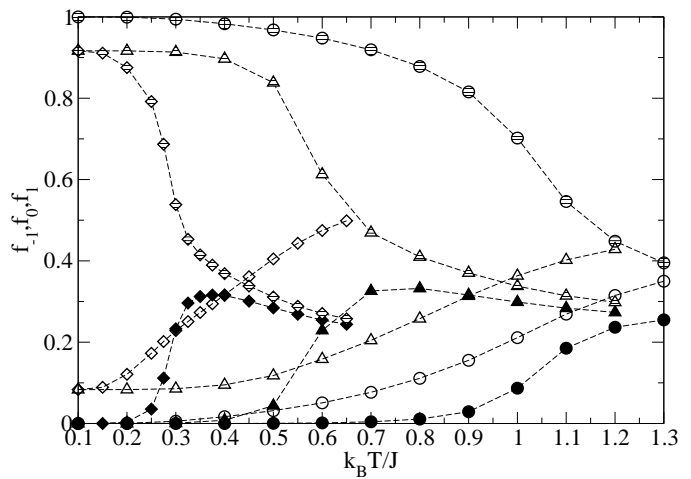}}
\caption{Fraction of spins in the state $n$, $f_n$, versus
temperature for lattices with $12\times 144$ spins at $(D/J,
\alpha)=(1.4, 0.4)$ (circles), $(1.7, 0.8)$ (triangles), and
$(1.98,0.95)$ (diamonds). The wetting transitions, in the three
cases, occur at $k_BT_w/J=$ 1.07, 0.60, and 0.315. Open, full, and
hatched symbols denote MC data for $n$=0, -1 and 1, respectively.
} \label{bcfigfrac}
\end{figure}

For critical wetting, one expects that the ``-1'' spins on the
left side boundary of the lattice will impose, at $T \ge T_w$, a
region of macroscopic extent, with, predominantly, ``-1'' spins in
the left part of the system. The wetting transition should be
indicated by anomalies and singularities in various
quantities~\cite{Diet}.

To identify and analyze the wetting transition, we performed MC
simulations for lattices with $N= L\times M$ sites or spins, see
figure~\ref{bcfiggeometry}. Critical properties of the transition
are believed to be affected by the two different correlation
lengths parallel and perpendicular to the interface between the
``-1'' and ``+1'' rich regions. The resulting anisotropic scaling
is taken into account by simulating lattices with $M \propto
L^2$~\cite{Alba}, i.e., with a constant generalized aspect ratio
$c=M/L^2$. The factor of $2$ corresponds to the ratio of the
critical exponents of the two correlation lengths~\cite{Alba}.

Specifically, we studied lattices with $c=1$, with  $L$ ranging
from 6 to 30, and $M$ ranging, accordingly, from 36 to 900. To
obtain data of the desired accuracy, runs with $10^7$ to $6 \times
10^7$ MC steps per spin were done. (Alternatively, cluster flip
MC algorithms, like the Swendsen-Wang or the Wolff algorithm ~\cite{SW}, may
be applied). To estimate error bars,
averages over runs with different random numbers were taken. In
fact, for the data depicted in the figures, error bars turned
out to be smaller than the symbol sizes, and are not shown.

\begin{figure}
\resizebox{1 \columnwidth}{!}{\includegraphics{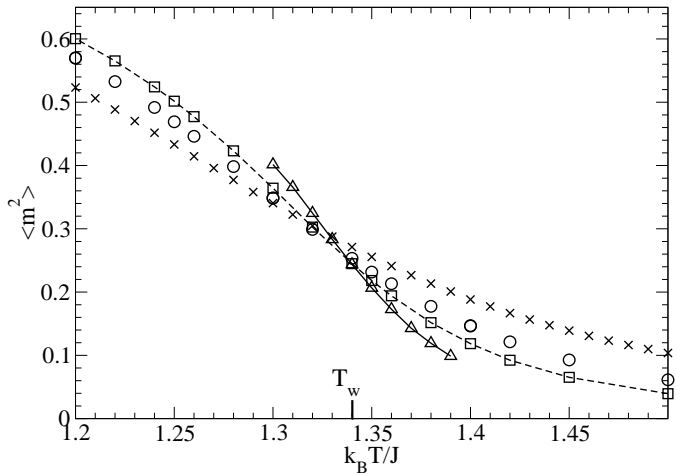}}
\caption{Second moment of the magnetization versus temperature at
$(D/J, \alpha)=(1.0, 0.2)$ for lattices with $(L,M)=(8,64)$
(crosses), $(12,144)$ (circles), $(16,256)$ (squares), and
$(24,576)$ (triangles) spins.} \label{bcfigm2}
\end{figure}

\begin{figure}
\resizebox{1 \columnwidth}{!}{\includegraphics{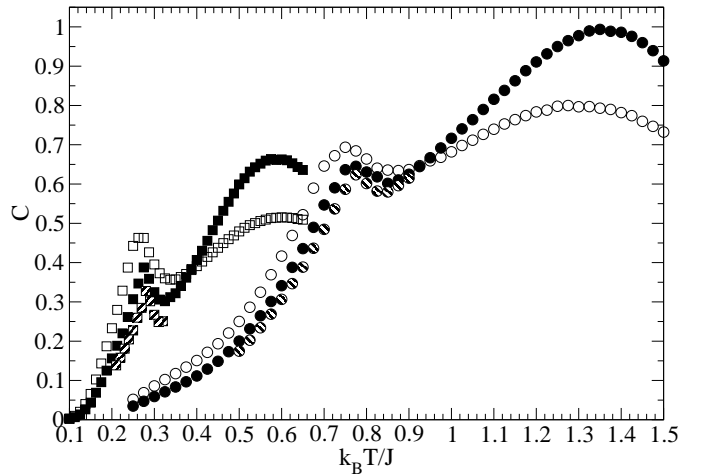}}
\caption{Specific heat $C$ as function of temperature for lattices
with $(L,M)=(8,64)$ (open symbols), $(12,144)$ (full), and
$(16,256)$ (hatched) spins for $(D/J, \alpha)=(1.0,0.8)$ (circles)
and $(1.98,0.95)$ (squares). Wetting occurs at $k_BT_w/J=$ 0.80
and 0.315, resp.} \label{bcfigsh}
\end{figure}

\begin{figure}
\resizebox{1 \columnwidth}{!}{\includegraphics{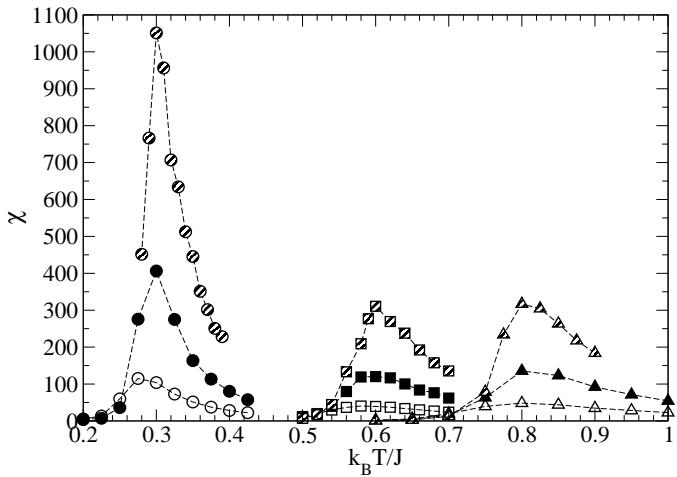}}
\caption{Susceptibility $\chi$, as function of temperature close
to wetting for lattices with linear size $M=64$ (open), $144$
(full), and $576$ (hatched) at $(D/J,\alpha)=(1.0, 0.8)$
(triangles), $(1.7, 0.8)$ (squares), and $(1.98, 0.95)$ (circles).
Wetting occurs at $k_BT_w/J=$ 0.80, 0.60, and 0.315, resp.}
\label{bcfigchi}
\end{figure}

\begin{figure}
\resizebox{1 \columnwidth}{!}{\includegraphics{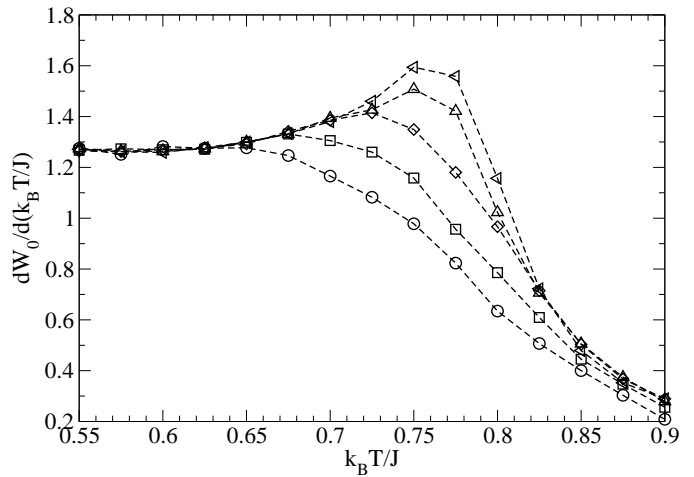}}
\caption{Temperature derivative of interfacial adsorption of ``0''
spins versus temperature at $D/J=1.0$ and $\alpha=0.8$ for
lattices with $(L,M)=(6,36)$ (circles), $(8,64)$ (squares),
$(12,144)$ (diamonds), $(16,256)$ (triangles up), and $(24,576)$
(triangles left) spins. Here, $k_BT_w/J$= 0.80.} \label{bcfigw0}
\end{figure}

The wetting transition is signalled by a variety of physical
quantities, as illustrated in
figures~\ref{bcfigfrac}-\ref{bcfigw0}. Examples are the specific
heat $C$, equation~(\ref{eq:C}), the susceptibility $\chi$,
equation~(\ref{eq:chi}), and the second moment of the
magnetization $\langle m^2 \rangle$.
Furthermore, we recorded the fraction of spins in the state $n$,
$f_{n}$
\begin{equation}
\label{eq:fn} f_n =\frac{1}{LM} \sum_{\langle (i,j)\rangle}
\delta_{S_{(i,j)},n}.
\end{equation}
Another interesting quantity is the interfacial
adsorption, $W_0$ \cite{Pesch}.
$W_0$ measures the surplus of ``0'' spins due to the interface
between the ``-1'' and ``+1'' rich regions induced by the
fixed boundary conditions:
\begin{equation}
\label{eq:W0} W_0=L(f_0 - F_0),
\end{equation}
where $F_0$ is the fraction of spins in state ``0'' when the fixed
boundary spins have always the same sign, say, ``+1''. To determine
$F_0$, separate MC runs, for lattices with the described boundary
conditions, were performed. The
behavior of the above quantities at the wetting transition will be
discussed in detail below. In addition, we recorded and monitored
several other quantities, as had been done before~\cite{Alba},
like magnetization histograms and profiles, as well as the Binder
cumulant~\cite{Bin}. Of course, in contrast to wetting induced by
surface fields~\cite{Alba}, the magnetization
histograms are no longer symmetric around zero magnetization due
to the reduced couplings at one of the boundaries. Finally, typical
MC equilibrium configurations are helpful in illustrating
wetting.

The fraction of spins in the state $n=0, \pm 1$, $f_{n}$, shows
the wetting transition in the, perhaps, clearest way. As seen from
figure~\ref{bcfigfrac}, $f_{-1}$ increases quite drastically close
to $T_w$, reflecting the spreading of the ``-1'' rich region from
the left boundary. This behavior holds at all values of $D/J$ and
$\alpha$ we studied. $f_{1}$ displays the corresponding inflection
point near $T_w$, now due to the quite drastic decrease in the
number of ``+1'' spins. Of course, there are finite-size effects
in $f_{1}$ and $f_{-1}$, with more pronounced changes approaching
$T_w$ for larger lattices.

It is worthwhile to take a closer look at $f_{0}$. As  seen in
figure~\ref{bcfigfrac}, one obtains $f_{0}(T=0)=1/L$ in the case
of pre-wetting at the left boundary by a single line of ``0''
spins, in particular for $D/J=1.7$ and $1.98$. There is no
pre-wetting in case of $D/J=1.4$ and $\alpha=0.4$, in accordance
with the considerations mentioned above, equation~(\ref{eq:pw}). For
$D/J=1.98$ and $\alpha=0.95$, $f_{0}$ is seen to grow rather
rapidly with increasing
temperatures. Inspection of typical MC equilibrium configurations
show an initial expansion of the ``0'' rich region at the left
boundary. However, eventually the ``0'' region will be bordered by
``+1'' as well as  ``-1'' domains, allowing for a meandering of
both interfaces. This ``entropic repulsion'' of the two interfaces
has been discussed before in the framework of critical interfacial
adsorption~\cite{Kroll}.

It is interesting to note that $f_{0}$, like $f_{-1(1)}$, exhibits
a, comparably weak, inflection point close to $T_w$. Again,
finite-size effects enhance the anomaly for larger lattices. We
shall return to the aspect when discussing $W_0$.

It has been suggested that, in simulations, the analysis of
$\langle m^2 \rangle$ may lead to reliable estimates of $T_w$ in
the two-dimensional BC model~\cite{Alba}. Like related quantities,
such as other moments of the magnetization and the Binder
cumulant, $\langle m^2 \rangle$ displays for two lattices of
different sizes an intersection point, approaching $T_w$ when
enlarging the lattices sizes. The rather small finite size effects and
the high statistical accuracy of MC
data for $\langle m^2 \rangle$  have been argued to be in
favor of using this quantity to determine $T_w$ ~\cite{Alba}. Indeed,
we confirmed the suggestion done for wetting imposed
by surface fields in our study
on wetting induced by fixed boundary spins.

Typical results are shown in figure~\ref{bcfigm2}, taking
$D/J=1.0$ and $\alpha=0.2$, with $L$ ranging from 6 to 24 (thence,
$M$ from 36 to 576). There are well-defined intersection points,
with only weak finite-size effects. The resulting estimates for
$T_w$ for this and other choices of $D/J$ and $\alpha$ are
depicted in figure~\ref{bcfigphadia}.

Our estimates for $T_w$, as obtained from the various quantities
discussed in this section, are, indeed, compatible with the
findings on the wetting transition temperature based on $\langle
m^2 \rangle$.

As exemplified in figure~\ref{bcfigsh}, the specific heat $C$ in
our MC study is found to display a maximum near the wetting
transition $T_w$. However, in contrast to the peak close to the
bulk transition, $T_c$, the height of the peak shrinks for larger
lattices. In fact, at $T_c$ one expects for sufficiently large
systems, a logarithmic divergence of the peak height for
continuous transitions in the universality class of the
two-dimensional Ising model. At the bulk transition of first-order,
the maximal specific heat is expected to diverge even more strongly
with a power-law in the number of sites. The lowering of the
height of the maximum,
when enlarging $L$, is consistent with the corresponding critical
exponent being negative. In fact, one expects a value of
$-2$~\cite{Alba}. The position of the maximum in $C$ is shifted
towards higher temperatures with increasing $L$, approaching
$T_w$.

Typical MC data close to $T_w$ for the susceptibility $\chi$ are
depicted in figure~\ref{bcfigchi}, for various values of $D/J$ and
$\alpha$. Obviously, $\chi$ exhibits  a pronounced maximum of
height $\chi^{\rm (max)}$, there, increasing rapidly with system
size, $L$. From analyzes of $\chi^{\rm (max)} (L) \propto
L^{\omega}$, we find that $\omega$ seems to approach $3$ for large
lattices, in agreement with the previous result~\cite{Alba}.

It is worth mentioning, that the maxima in $C$ and  $\chi$ may
be useful to show the ending of the wetting line, $T_w$, in the bulk
transition line, $T_c$, see figure~\ref{bcfigphadia}. For
instance, at $D/J=1.98$ and $\alpha \le 0.6$, we observe that
the specific heat $C$ displays a unique maximum close to the
bulk transition, whose height increases
significantly more strongly than logarithmically with $L$. This behavior
indicates the bulk transition of first-order. Likewise, $\chi$
exhibits a unique maximum, with the size-dependent effective exponent
$\omega$ increasing with $L$, up to about 2 for the sizes we studied.

Finally, let us discuss the MC findings on the interfacial
adsorption $W_0$, equation~(\ref{eq:W0}), near the wetting
transition. As mentioned above, the fraction of spins in state
``0'', $f_{0}$, displays an inflection point near $T_w$. With the
interface between the ``-1'' and ``+1'' rich regions moving
towards the center of the lattice, the interface may fluctuate or
meander more strongly, and the number of ``0'' spins may grow more
rapidly. To disentangle bulk and interface effects on the ``0''
spins, we consider $W_0$, measuring, roughly, the width of the
(fictitious) stripe of ``0'' spins due to the
interface~\cite{Pesch}.

As depicted in figure~\ref{bcfigw0} for the example $D/J=1.0$ and
$\alpha=0.8$, the temperature derivative of $W_0$,
$dW_0/d(k_BT/J)$, shows, in fact, a clear maximum near $T_w$. For
larger $L$, the position of the maximum moves towards $T_w$, with
increasing height. The MC data seem to be consistent with a
singularity at $T_w$ in the thermodynamic limit. The height of the
maximum may diverge, with $L$, either logarithmically or in form
of a power law with a small exponent, $\le 0.2$. We observed
similar features for other choices of $D/J$ and $\alpha$ as well. The
detailed analysis is desirable, but it would require much more
computational efforts. Accordingly, it is beyond the scope of the
present study.

We note in passing that we found the interfacial adsorption $W_0$
at the tricritical point to increase, for quadratic lattices, as
$W_0(T_t) \propto L^a$, with $a$ being close to 4/5, $L$ ranging
from 40 to 100. The finding
confirms a previous scaling argument~\cite{Kroll}.

\section{Summary}
\label{sec:summary}

We studied the Blume-Capel model on the square lattice, with
nearest neighbor exchange interactions, $J$, and the single-ion
anisotropy term, $D$. We used, mainly, Monte Carlo simulations,
both of Metropolis and Wang-Landau type. For selected cases of
$D/J$, we determined accurately the bulk phase transition
temperature, $T_c$, by applying a modified Wang-Landau scheme.

Wetting has been induced by appropriate boundary conditions. Spins
at two opposite boundaries are fixed in the different states
``-1'' and ``+1''. Moreover, at one of the boundaries, say, the
one with ``-1'' spins, the exchange interaction is reduced by the
factor $\alpha$, where $0<\alpha<1$.

Wetting lines, $T_w$, have been determined for various values of
$D/J$, yielding continuous or bulk transitions of first-order. To
estimate $T_w$, the second moment of the magnetization, $\langle
m^2 \rangle$ turned out to be a useful quantity, showing only weak
finite-size effects. The spreading of the ``-1'' rich region near
$T_w$ may be clearly monitored by the fraction of spins in state
``-1'', $f_{-1}$. The ``0'' spins may lead to pre-wetting next to
the ``-1'' boundary depending on $D/J$ and $\alpha$.

For continuous bulk transition, the wetting line, at fixed
$D/J$, $T_{w}(\alpha)$ is found to decrease monotonically with the
reduction factor $\alpha$, where $T_{w}(0)= T_c$ and $T_{w}(1)=0$.
In the case of a bulk transition of first-order, we observe that
the wetting line ends at the bulk transition line at a non-zero
value of $\alpha$. Especially, for $D/J=1.98$, the merging seems
to occur close to $\alpha=0.6$.

Critical properties at the wetting agree with those obtained recently
for wetting imposed by surface fields, including
anisotropic scaling and critical exponents of the susceptibility
and specific heat.

Last, but not least, we monitored the adsorption of ``0'' spins at the
interface between
``-1'' and ``+1'' rich regions. Due to the strong meandering of
the interface, the adsorption, $W_0$, grows rapidly near $T_w$,
associated, possibly, with a singularity in the temperature
derivative of $W_0$.

\begin{acknowledgement}
Our study has been motivated by an inspiring talk of Kurt Binder
on wetting. It is a pleasure to thank him for a very useful
correspondence on this topic as well. N.G. Fytas is grateful to
the Departamento de F\'{i}sica Te\'{o}rica I, Universidad
Complutense de Madrid (Spain), for providing access to computing
resources, where part of the current simulations has been
performed.
\end{acknowledgement}

\end{document}